\documentstyle[a4,11pt,epsf]{article}
\newcommand{\be}{\begin{eqnarray}}
\newcommand{\ee}{\end{eqnarray}}
\newcommand{\nn}{\nonumber}
\newcommand{\ovl}{\overline}
\newcommand{\ra}{\rightarrow}

{\newcommand{\lsim}{\mbox{\raisebox{-.6ex}{~$\stackrel{<}{\sim}$~}}}
{\newcommand{\gsim}{\mbox{\raisebox{-.6ex}{~$\stackrel{>}{\sim}$~}}}

\def\0n{0\nu\beta\beta}

\def\Lbar{L\hskip-1.5mm/}

\begin{document}
\begin{center}
{\Large \bf Light Lepton Number Violating Sneutrinos and the Baryon 
Number of the Universe} \\
\vspace{0.5cm}
{\large H.V. Klapdor-Kleingrothaus $^{\dagger}$, 
St. Kolb$^{\dagger,\ddagger}$ and 
V.A. Kuzmin$^{\ddagger}$}\\
\vspace{0.3cm}
{$^{\dagger}$ Max-Planck-Institut f\"ur Kernphysik,
P.O. 10 39 80, D-69029 Heidelberg, Germany} \\ \vspace{0.3cm}
{$^{\ddagger}$ Institute for Nuclear Research of the Russian Academy
of Sciences, 60th October Anniversary Prospect 7a, 117321 Moscow, Russia}
\end{center}

\vspace{0.5cm}

\begin{abstract}
Recent results of neutrino oscillation experiments point to a non-vanishing
neutrino mass. Neutrino mass models favour Majorana-type neutrinos. In such
circumstances it is natural that the supersymmetric counterpart of the 
neutrino, the sneutrino, bears also lepton number violating properties.
If the amount of lepton number violation is large enough the sneutrino
may be the Cold Dark Matter in the universe. On the other hand, the fact 
that the universe exhibits an asymmetry in the baryon and antibaryon 
numbers poses constraints on the extent of
lepton number violation in the light sneutrino sector if the electroweak
phase transition is second or weak first order. From the requirement that 
the Baryon Asymmetry of the Universe should not be washed out by sneutrino 
induced lepton number violating interactions and sphalerons
below the critical temperature of the electroweak phase transition we find 
that the mass-splitting of the light sneutrino mass states is compatible
with the sneutrino Cold Dark Matter hypothesis only for heavy gauginos
$M_1,M_2\gsim 500 GeV$ and opposite sign gaugino mass parameters.
 
%
\end{abstract}

\vspace{0.5cm}
\noindent
{\bf I \hspace{0.2cm} Introduction}\\

There are hints from neutrino oscillation experiments that the neutrino 
is massive (\cite{kamiokande} and refs. therein. For a recent overview 
see {\it e.g.} \cite{oscover}). In 
most neutrino mass models the neutrino is of Majorana-type, {\it i.e.} it
violates lepton number $L$. If this is indeed the case the next generation
of experiments searching for neutrinoless double beta ($\0n$) decay, which 
are the only experiments capable of deciding on the nature of the 
neutrino, possibly will be able to indeed observe a $\0n$-decay
signal (for a recent overview see {\it e.g.} \cite{klaptrento}).

On the other hand, it has been shown in \cite{theorem} that if the 
neutrino is a massive 
Majorana field the low energy effective theory of the supersymmetric 
extension of the Standard Model (for a phenomenological overview
see {\it e.g.} \cite{haberkane}) will contain mass terms 
for the sneutrino
which violate $L$ too, regardless of the mechanism which is
responsible for the generation of sneutrino masses in the unbroken 
theory. 
In \cite{susyseesaw} a model containing heavy $SU(2)$ singlet 
sneutrino fields and $L$-violating mass terms involving these 
fields has been examined (such models have 
been considered previously in connection with the generation of the 
Baryon Asymmetry of the Universe (BAU) at 
some high temperature, see {\it e.g.} \cite{baugen}). In both cases
below the electroweak symmetry breaking scale 
the weak states $\tilde{\nu},\tilde{\nu}^*$ are no longer mass
states and the resulting mass states violate $L$, exhibit a mass-splitting 
and give rise to $L$-violating processes  
which have been analyzed {\it e.g.} for the Next Linear Collider in 
\cite{phen} and for $\0n$ decay in \cite{sndb}.

It has been pointed out in \cite{cdm} that in a scenario where the
light sneutrino mass states exhibit a mass-splitting
the lightest sneutrino could account for the Cold Dark Matter (CDM) in the 
universe. This is due to the fact that the sneutrino mass states couple 
``off-diagonally'' to the $Z^0$ on grounds of Bose statistics and
angular momentum conservation so that the arguments excluding ordinary
sneutrinos from constituting a substantial fraction of the CDM are
not valid in the case of light sneutrinos exhibiting a mass difference.
However, it has been shown in \cite{cdm} that the mass difference
should be of order ${\cal O}(\mbox{few}\; GeV)$.

On the other hand sneutrino mediated $L$-violating reactions, like 
any other $L$-violating process, may be dangerous for the BAU  
due to sphalerons \cite{sph}. The constraints on the 
$L$-violating sneutrino properties stemming from the requirement that a BAU 
generated at some early epoch in the evolution of the universe 
should not be destroyed by sneutrino-induced interactions during a
later epoch are the subject of this note. 

\vspace{0.5cm}
\noindent
{\bf II \hspace{0.2cm} Lepton number violation in the sneutrino sector}\\

In the following the discussion will be restricted to the one-generation
case. We plan to investigate the impact of possible $CP$ violation in a
multi-generation scheme for the light sneutrino sector elsewhere.  
It has been pointed out in \cite{theorem} that the low energy ({\it i.e}
below the scale where the $SU(2)_L \times U(1)_Y$ gauge group is broken
into $U(1)_{em}$) sneutrino mass terms can be written as
\be \label{effective}
{\cal L}_{mass}^{eff} = 
-\frac{1}{2}(\tilde{m}_D^2 \tilde{\nu}_L \tilde{\nu}_L^* + 
             \tilde{m}_M^2 \tilde{\nu}_L \tilde{\nu}_L + h.c.).
\ee
Here $\tilde{m}_D^2$ contains as usual contributions from $V_{soft}$
and from the D-term, whereas $\tilde{m}_M^2$
violates $L$ explicitly and may have its origin at some high energy scale. 
Furthermore, loops containing Majorana neutrinos and neutralinos 
induce contributions to $\tilde{m}_M^2$ radiatively. Expression 
(\ref{effective}) is valid if no right-handed sneutrino sector
is present in the theory and independent of the mechanism which generates
the $L$-violating sneutrino mass. Writing 
$\tilde{\nu}_L=1/\sqrt{2}(\tilde{\nu}_1 + i \tilde{\nu}_2)$
where $\tilde{\nu}_{1,2}$ are real, the resulting mass states are simply
$\tilde{\nu}_{1,2}$ with masses $m_{1/2}^2=\tilde{m}_D^2 \pm 
\tilde{m}_M^2$ and the mass difference is 
$\Delta m^2 = 2 \tilde{m}_M^2$.

If a right-handed sector is included
the sneutrino mass terms after electroweak symmetry breaking 
without an explicit $L$-violating mass term for the $SU(2)$ doublet 
sneutrinos are
\cite{susyseesaw}
\be \label{massmatrix}
{\cal L}_{mass}=-\frac{1}{2} (\phi_1,\phi_2) 
\left( 
\begin{array}{cc} {\cal M}_+^2 & 0 \\
                                0     & {\cal M}_-^2       
\end{array}
\right)
\left( \begin{array}{c} \phi_1 \\ \phi_2 \end{array} \right)
\ee
where $\phi=(\tilde{\nu}_i,\tilde{N}_i)^T$ and 
$\tilde{N}=(\tilde{N}_1 + i \tilde{N}_2)/\sqrt{2}$ is the $SU(2)_L$
singlet field. The matrices ${\cal M}_{\pm}^2$ are defined as 
\be  
{\cal M}_{\pm}^2 =
\left( \begin{array}{cc}
m^2_{\tilde{L}} + \frac{1}{2}m^2_Z \cos 2 \beta + m_D^2 & 
m_D (A_{\nu} - \mu \cot \beta \pm M) \\
m_D (A_{\nu} - \mu \cot \beta \pm M) &
M^2 + m_D^2 + m_{\tilde{N}}^2 \pm 2 B_N M \end{array} \right) \; .
\ee
The parameters $m_{\tilde{N}}^2$ and $A_{\nu}$ are
contained in $V_{soft}$ in analogy to the charged sfermion sectors.
Furthermore $V_{soft}$ contains a term $M B_N \tilde{N} \tilde{N} + h.c.$
which violates $L$. The remaining entries of the mass matrix 
(\ref{massmatrix}) are the
$F$-terms stemming from the superpotential which in comparison to the 
MSSM contains an additional term  $M \hat{N} \hat{N}$. 
The Dirac neutrino mass is $m_D=\lambda v_2$,
$\lambda$ being a Yukawa coupling, $v_i/\sqrt{2}$ is the vacuum 
expectation value of the neutral component of the $SU(2)$ doublet Higgs 
$H_i$, and $\tan \beta=v_2/v_1$. It is natural that $M \gg m_Z$
since $m_{\nu} \approx m_D^2/M$, 
$\mu,A_{\mu},m_{\tilde{L}} \sim {\cal O}(m_Z)$ since sfermions
with nontrivial $SU(2)_L \times U(1)_Y$ transformation properties
should not be much heavier than $100GeV$-$1TeV$, 
whereas $B_N,m_{\tilde{N}}$ may a priori be much bigger since they
pertain to the $SU(2)_L \times U(1)_Y$ singlet field $\tilde{N}$,
{\it cf.} the discussion in \cite{susyseesaw}. 

For the matrix (\ref{massmatrix}) the mass states read
\be \label{states}
\xi^l_{1,2} = \cos \Theta_{\pm} \tilde{\nu}_{1,2} + 
              \sin \Theta_{\pm} \tilde{N}_{1,2} \nn \\
\xi^h_{1,2} = - \sin \Theta_{\pm} \tilde{\nu}_{1,2} +
              \cos \Theta_{\pm} \tilde{N}_{1,2} 
\ee
where the angles $\Theta_{\pm}$ diagonalize ${\cal M}^2_{\pm}$
and the indices $l,h$ refer to light and heavy sneutrino mass states.
In leading order in $1/M$ the mass difference of the light sneutrino 
states is
\be \label{difference}
\Delta m^2 = m^2_{\xi_2^l} - m^2_{\xi_1^l} = 
4 \frac{m_D^2}{M} (A_{\nu} - \mu \cot \beta - B_N)
\ee
in accordance with \cite{susyseesaw}. 

In the following it will not be distinguished further between the 
models eq. (\ref{effective}) and eq. (\ref{massmatrix}) and the light
sneutrino mass states will be denoted by $\xi^l_{1,2}$. Note that
the mass difference $\Delta m$ of the light states is related to 
$\Delta m^2$ by 
\be \label{relation}
\Delta m=m_{\xi^l_2}-m_{\xi^l_1}=\Delta m^2/(m_{\xi^l_1}+m_{\xi^l_2}) \ .
\ee

\vspace{0.5cm}
\noindent
{\bf \hspace{0.2cm} III Baryon number depletion by sneutrinos below 
the critical temper-\-
\hspace*{1cm} ature}\\

In both scenarios eqs. (\ref{effective}),(\ref{massmatrix}) sneutrinos 
give rise to $L$-violating processes,
{\it e.g.} sneutrino decays or 2$\leftrightarrow$2 scatterings.
These scatterings have the potential to erase an asymmetry in the number 
of baryons and
antibaryons in the early universe. Such an asymmetry has to be generated
somewhen during the evolution of the universe in order to  
explain the absence of antimatter in the universe observed (for an 
overview see {\it e.g.} \cite{kotu}). In the following it is assumed
that at some high temperature above the electroweak symmetry breaking
scale the BAU has been generated by
some mechanism in the right amount. One example of such a mechanism is
the decay of heavy right-handed neutrinos and/or sneutrinos which 
incorporates the three basic conditions for the generation of the BAU:
baryon or lepton number violation, $CP$-violation and out of thermal 
equilibrium circumstances. All three conditions may be satisfied by 
the decay of
$SU(2)_L$ singlet sneutrinos, in particular the out of equilibrium condition
requires them to be very massive (${\cal O}(10^{10}GeV)$) in accordance
with the considerations mentioned above ($M \gg {\cal O}(m_Z)$).

However, at temperatures below the electroweak symmetry breaking scale 
$T \sim {\cal O}(100GeV)$ the light sneutrino states violate $L$ too 
and sneutrino interactions 
may bring the distributions of leptons and antileptons
into equilibrium. An estimation of interactions induced by Majorana 
neutrinos below $T_C$ has been given in \cite{utpal}.
The approximate criterium for equilibration is that
the rate of the process in question should be bigger than the expansion
rate of the universe $H(T)=1.7 \sqrt{g_*}\;T^2/M_{Pl}$ (in a radiation 
dominated universe) where $g_*\approx 200$ in the MSSM and 
$M_{Pl} \approx 10^{19} GeV$.

As long as sphaleron-mediated processes are operative the lepton and 
baryon numbers are both proportional to the combination $B-L$ \cite{sph}.
As a consequence the asymmetry in baryons vanishes if the asymmetry
in leptons is somehow erased. Hence, if there is a temperature range
between the critical temperature of the $SU(2)_L \times
U(1)_Y \ra U(1)_{em}$ phase transition and the temperature $T_{out}$
at which sphaleron-mediated processes drop out of equilibrium ({\it i.e.} 
$\Gamma_{Sph}(T_{out}) < H(T_{out})$ where $\Gamma_{Sph}(T)$ is the sphaleron 
rate) so that
\be \label{range}
T_{out} < T < T_C
\ee
then $L$-violating interactions are potentially dangerous for the BAU. 
The presence of a temperature range satisfying (\ref{range}) is not
possible in models where the BAU is generated during the electroweak
phase transition itself (Electroweak Baryogenesis) since in such models 
the sphaleron-mediated processes must be switched off immediately below
$T_C$. Therefore any constraints on the $L$-violating properties of light
sneutrinos do {\it a priori} not hold in the context of Electroweak 
Baryogenesis or any other model where sphaleron-mediated processes are out
of equilibrium immediately after the electroweak phase transition.

In order to estimate the temperature range (\ref{range})
the following assumptions are made: 
the evolution of the vacuum expectation value ({\it vev}) of the Higgs 
fields $\langle v (T)\rangle$ 
is described by
\be \label{vacvalue}
\langle v (T) \rangle = \langle v (0) \rangle (1 - T^2/T_C^2)^{1/2}\;\;,
\langle v (0) \rangle = 246 GeV\;.
\ee
This behaviour of the {\it vev} is valid for a second order phase 
transition and approximately valid for a weak first order phase 
transition. The exact behaviour of $\langle v (T) \rangle$  
depends on the SUSY parameters, but for our phenomenological
purposes we will satisfy ourselves with eq. (\ref{vacvalue}). Lattice
simulations suggest that $T_C \approx 150 GeV$ for a Higgs mass of
$70 GeV$ \cite{valuetc} and that it should be higher for larger values of the
Higgs mass, but in order to keep the results general we will vary $T_C$ 
freely between $50 GeV$ and $250GeV$. In models of electroweak
baryogenesis sphalerons have to drop out of equilibrium immediately
after the phase transition which translates into the condition 
\cite{sphrate} $\langle v (T=T_C) \rangle / T_C > 1$ in contrast to 
(\ref{vacvalue}), see the comment above. 

\begin{figure}[t]
\vspace*{-1cm}
\hspace*{3cm}
\epsfxsize65mm
\epsfbox{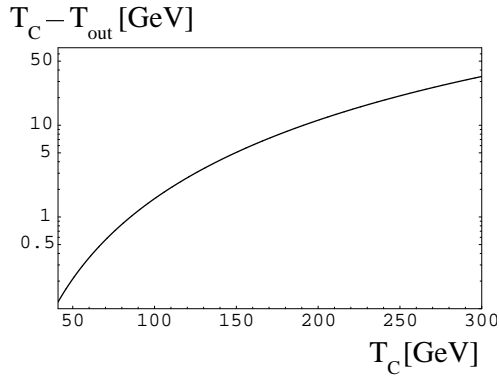}
\vspace{-0.5cm}
\caption{The difference of the critical temperature $T_C$ and 
the sphaleron 
freezing out temperature $T_{out}$ determined by 
$\Gamma_{sph}(T_{out})=H(T_{out})$ in dependence of $T_C$.}
\label{tempdiff}
\end{figure}

The sphaleron rate below $T_C$
is described by (\cite{sphrate} and refs. therein)
\be \label{sphrate}
\Gamma_{Sph} \approx 2.8 \cdot 10^5 \; T^4 \; \kappa \; 
            \left(\frac{\alpha_W}{4 \pi}\right)^4
            \left(\frac{2 m_W (T)}{\alpha_W T}\right)^7
            \exp\left(-\frac{E_{sp}(T)}{T}\right)
\ee
where 
\be
m_W (T) = \frac{1}{2} g_2 \langle v (T) \rangle \ ,
\ee
the free energy of the sphaleron configuration is given by
\be
E_{Sph} (T) = \frac{2 m_W (T)}{\alpha_W} B \left(\frac{m_H}{m_W}\right),
\ee
%
%
%
$B(0)=1.52,\; B(\infty)=2.72$ and $\kappa$=exp(-3.6) \cite{moore}.
In figure \ref{tempdiff} the temperature range $T_C-T_{out}$ is
plotted as a function of $T_C$. It becomes smaller than one
for small values of $T_C \lsim 100 GeV$ but is of order 
${\cal O}(10 GeV)$ for $T_C \sim 200 GeV$. 

For $L$-violating sneutrino decays and $L$-violating sneutrino 
mediated scatterings
the damping of the preexisting Lepton number (per comoving volume)
$L_i$ is described by the Boltzmann equation for the evolution
of $L$ and is given by (see {\it e.g.} \cite{kotu})
\be \label{evolution}
L(z)=L_i \exp \left[ 
      - \int\limits_{z_c}^{z_{out}} d z' z' 
                 [g_* \frac{n_{\tilde{\nu}}}{s} 
                  \Gamma_D(z')+ n_{\gamma} 
                  \langle \sigma |v| \rangle)]/H(T=m_{\tilde{\nu}}) \right]
\ee
where $z=m_{\tilde{\nu}}/T$, $\Gamma_D$ is the $L$-violating sneutrino 
decay width, $\langle \sigma |v| \rangle$ is the thermally averaged
cross-section of sneutrino mediated $L$-violating scatterings and $n_i$
is the number density of particle $i$ (in the scattering term the
number density of the initial state particles has been assumed to be the
photon number density). In order to derive bounds
on $L$ violation we will allow that during
the epoch $T_{out}<T<T_C$ $L$ may be depleted by a factor $k < 1$,
that is we assume the preliminary $L$ density to be 
$n_L/n_{\gamma}\approx 10^{-10}/k$. 
%


\vspace{0.5cm}
\noindent
{\bf IV \hspace{0.2cm} Constraints on the sneutrino mass-splitting}\\

In the following two different cases will be distinguished: 
{\bf I)} 
at least one gaugino (see below) or slepton other than the sneutrino
are lighter than $T_C$, that is the number density of particles 
satisfying this condition is approximately the photon number density;
{\bf II)} all gauginos and sleptons besides the sneutrino are 
(substantially) heavier than $T_C$ so that the number density of these
particles is exponentially suppressed and one sneutrino is the lightest 
supersymmetric particle (LSP). The discussion will be focused on the
case of sneutrino CDM, that is on a mass difference of order $5GeV$
and a sneutrino LSP mass of $m_{LSP}\sim70 GeV$ \cite{cdm}.  

\begin{figure}[t]
\hspace*{0.3cm}
\epsfxsize130mm
\epsfbox{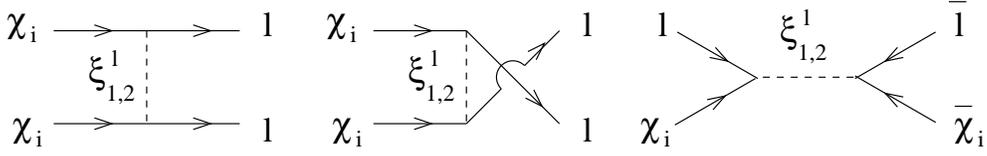}
\caption{Graphs for $2\leftrightarrow2$ scatterings of gauginos 
$\chi_i$ and leptons $l$
mediated by the light sneutrino states $\xi^l_{1,2}$.} 
\label{annihilation}
\end{figure}

{\bf I)} In order to keep the results as general as possible only 
scatterings will be taken into account in what follows since these
depend to a smaller extent on the exact mass relations of the
particles involved than $L$-violating decays do. Then relevant processes
capable of depleting $L$ are $2 \leftrightarrow 2$ scatterings as for 
example scattering of neutralinos or charginos into leptons mediated
by $L$-violating sneutrinos (see figure \ref{annihilation}) 
\be \label{reaction}
\chi_i \chi_j \longleftrightarrow l_l l_m \ \ , \ \ 
\chi_i l_l \longleftrightarrow \ovl{\chi}_j \ovl{l}_m 
\ee
where $\chi_k$ represents any of the chargino or neutralino
mass fields present in the plasma at $T<T_C$ 
and $l_n$ is a charged or uncharged lepton. In the one generation case 
$l$=$m$. However, since the
entries of the neutralino mixing matrix connecting gauginos with
higgsinos are proportional to the $vev$ this mixing will be 
suppressed by $\langle v(T=T_{out})\rangle/\langle v(T=0)\rangle$ 
and therefore $\chi$ will
be assumed to be $\tilde{B}$ with associated mass $M_1$ and
$\tilde{W}^3,\tilde{W}^{\pm}$ with associated mass $M_2$.

The following discussion will be restricted to the process eq.
(\ref{reaction}), but for other examples as 
$\tilde{l}^{\pm}\tilde{l}^{\pm} \longleftrightarrow W^{\pm} W^{\pm}$ 
the same arguments hold.
The constraint on the mass-splitting follows from the relation
\be \label{boundeq}
\int\limits_{z_C}^{z_{out}} d z' z' \frac{1}{H(m_{\tilde{\nu}})}
                  n_{\gamma} \langle \sigma |v| \rangle > \ln k \ .
\ee
The zero-temperature cross-section of
this reaction is proportional to $(\Delta m^2)^2$ since \cite{theorem}
\be \label{propagator}
\langle 0|T[\tilde{\nu}(x) \tilde{\nu}(y)]|0 \rangle &=&
\Delta m^2 \frac{i}{2} \int \frac{d^4 k}{(2 \pi)^4} 
\frac{e^{-ik(x-y)}}
     {(m^2_{\xi^l_1} - k^2)(m^2_{\xi^l_2} - k^2) + i \epsilon}
\ee
where for model eq. (\ref{massmatrix}) the contributions of the heavy
states have been neglected. Therefore
the final bound on $\Delta m$ depends only weakly on the precise
value of $\langle \sigma |v| \rangle$ (and on $k$) and for our 
purposes it is sufficient to approximate the thermally averaged 
scattering cross section by
\be \label{twotwocross}
\langle \sigma |v| \rangle \approx 
(\Delta m^2(T))^2 \frac{\alpha_W^2 T^2}{(T^2+m_{\tilde{\nu}}^2)^4} \ \ ,
\ee
where $\alpha_W$ is the weak coupling constant and for non-vanishing
temperatures and a second order phase transition (see eq. \ref{relation})
\be
\frac{\Delta m (T)}{\Delta m (T=0)}= 
\frac{\Delta m^2 (T)}{\Delta m^2 (T=0)}=
\frac{\langle v (T)\rangle^2}{\langle v (T=0) \rangle^2}=
\left( 1 - \frac{T^2}{T_C^2} \right) \ .
\ee
%
%
The full zero temperature cross section can be found in \cite{phen}. 

\begin{figure}[t]
\hspace*{3mm}
\epsfxsize130mm
\epsfbox{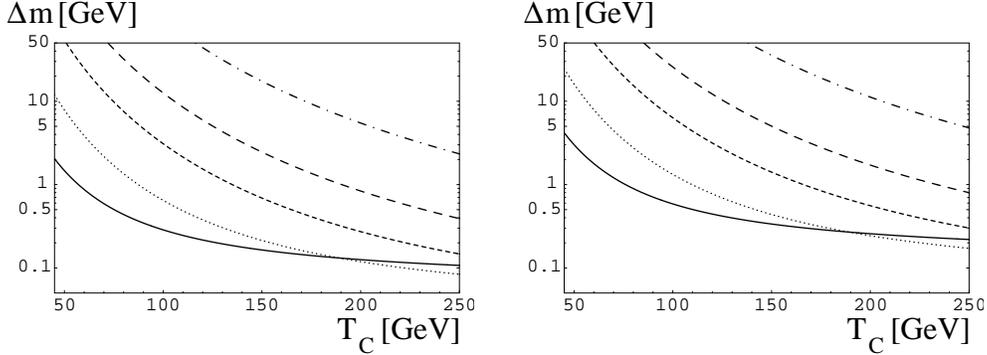}
\vspace{-0.5cm}
\caption{The limits on the zero temperature light sneutrino mass-splitting
splitting $\Delta m (T=0)$ in case {\bf I)} (see text) in dependence of 
the critical temperature $T_C$ from eq. (\ref{boundeq}) (the regions 
above the curves are excluded) 
for $k=0.3$ (left) and $k=0.01$ (right) and for different values of 
the sneutrino mass: $m_{\xi^l}=70 GeV$ (solid curve),
$m_{\xi^l}=150 GeV$ (dotted curve), $m_{\xi^l}=300 GeV$
(short-dashed curve), $m_{\xi^l}=500 GeV$ (long-dashed
curve), $m_{\xi^l}=1TeV$ (dot-dashed curve).} 
\label{figresult}
\end{figure}

%
The results, which hold for any generation, are plotted in figure 
\ref{figresult} for several values for the sneutrino mass and $k$
in dependence of $T_C$ which has been varied freely between $50GeV$
and $250GeV$. For example for $T_C \approx 150 GeV$ the limits range 
from $\Delta m \lsim {\cal O}(few \ 100 MeV)$ to 
$\Delta m \lsim {\cal O}(few \ GeV)$ for the chosen sneutrino masses.
The limit on the mass-splitting becomes less stringent
for smaller values of $T_C$ (since the temperature range 
(\ref{range}) becomes smaller) and for larger values of 
$m_{\tilde{\nu}}$. For small values of $T_C$ and large values of 
$m_{\tilde{\nu}}$ the mass-splitting is basically not constrained.

{\bf II)} In the case that all gauginos and sleptons but the
sneutrinos have already decayed away at temperatures above $T_C$
the processes depleting $L$ are $2\leftrightarrow2$ scatterings 
of sneutrinos into neutrinos and $L$-violating decays of non-LSP 
sneutrinos into the LSP sneutrino $\xi^h \ra \xi^{LSP}\nu\nu$ mediated
by neutralinos. 

\begin{figure}[t]
\epsfxsize130mm
\hspace*{0.3cm}
\epsfbox{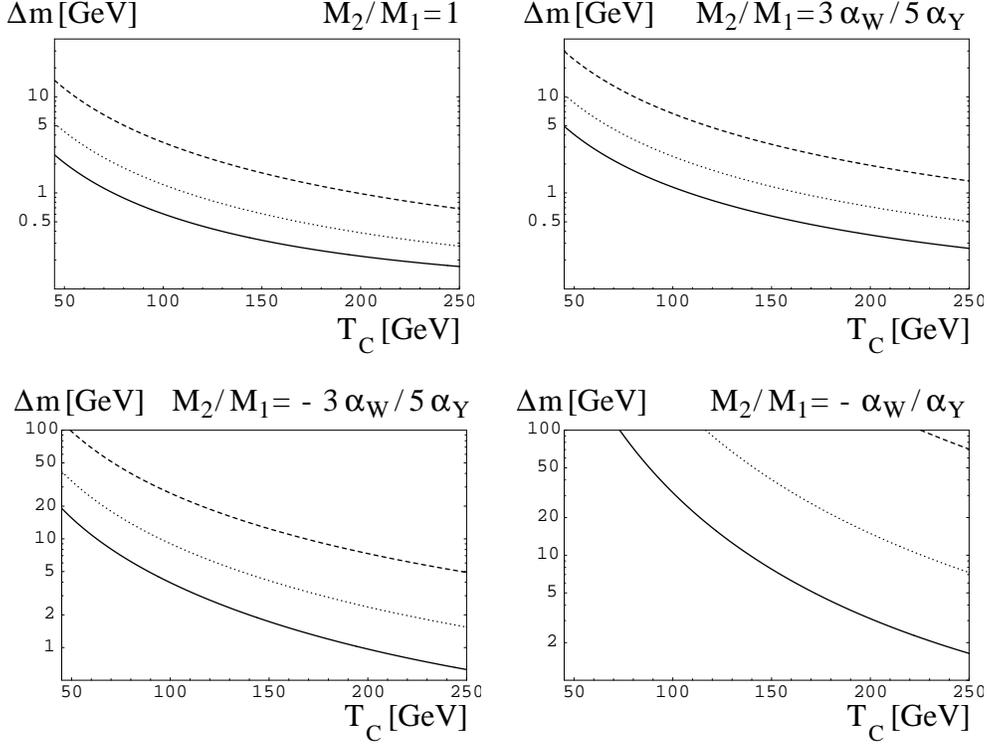}
\vspace{-0.8cm}
\caption{The limits on $\Delta m(T=0)$ from eqs. 
(\ref{sneuinitial}),(\ref{boundeq}) in case {\bf II)} (see text)
for several values of the
ratio $M_2/M_1$ in dependence on the critical temperature and
for a depletion factor $k=1/3$. The regions above the lines are 
excluded. The limits correspond to 
the parameters $M_1=300 GeV$ (solid line), $M_1=500 GeV$ (dotted line)
and $M_1=1 TeV$ (dashed line).}
\label{ressneuinitial}
\end{figure}

The cross-section for gaugino-mediated $L$-violating scattering of 
sneutrinos into neutrinos can be approximated by (compare with 
\cite{cdm})
\be \label{sneuinitial}
\langle \sigma |v| \rangle \approx 
\frac{(\Delta m^2 (T))^2}{T^4} 
\left( \frac{\alpha_Y M_1}{T^2+M_1^2} +
       \frac{\alpha_W M_2}{T^2+M_2^2} \right)^2
\ee
where $\alpha_Y$ is the coupling associated with $U(1)_Y$.  
This cross-section depends sensitively on the relation $M_2/M_1$.
In figure \ref{ressneuinitial} the bounds on the $\Delta m$ are
shown for several values of $M_1$ and for the cases of $M_2/M_1$
considered in \cite{cdm,newpap} which have been shown to be compatible
with the sneutrino CDM hypothesis as long as $M_1 \gsim 200 GeV$
(see next section).

The resulting limits are more stringent than the bounds in {\bf I)}
if $M_1$ and $M_2$ have the same sign. This is simply due to the
propagator structure of the two different scattering processes. 
Even for very large gaugino masses of order ${\cal O}(TeV)$ the
mass-splitting $\Delta m (T=0)$ is only of order ${\cal O}(few \ GeV)$
for plausible values of $T_C$. This statement is also valid in the limit
$M_2 \ra 0$. On the other hand, if $M_1$ and 
$M_2$ are of opposite sign there is a partial cancellation of the
contributions and the resulting limits on $\Delta m (T=0)$ become
weaker. In particular for $T_C \approx 200 GeV$ for 
$M_2/M_1=-3\alpha_W/5\alpha_Y$ ($-\alpha_W/\alpha_Y$) a mass difference
$\Delta m \sim {\cal O}(few \ GeV)$ is allowed if $M_1 \approx 500 GeV$
($300 GeV$).

The decays of non-LSP sneutrinos into the LSP sneutrino have been
discussed in detail in \cite{newpap}. Neglecting gaugino mixing
the zero-temperature decay width of the $L$-violating mode to 
leading order in the parameter $\Delta m^2$ is
\be
\Gamma_{\Lbar}\approx \frac{\alpha_W^2}{16 \pi}
    \frac{(\Delta m^2(T))^5}{(m_{\chi}^2-m_+^2)^4 \sqrt{m_+^2}} 
    \sim (\Delta m^2 (T=0))^5\left(1-\frac{T^2}{T_C^2}\right)^5
\ee   
where $m_+^2=m_{\xi^{LSP}}^2+m_{\xi^h}^2$. This expression does not
result into a constraint when inserted into eq. (\ref{evolution}) due
to its dependence on the fifth power of the suppression factor 
$(1-T^2/T_C^2)$: {\it e.g.} for $T_C \lsim 250 GeV$ the constraint
on the mass-splitting is $\Delta m \lsim {\cal O}(100 GeV)$ and for
smaller values of $T_C$ becomes even less stringent. 


\vspace{0.5cm}
\noindent
{\bf V \hspace{0.2cm} Comparison with low-energy constraints and 
implications for sneu-\-
\hspace*{0.8cm} trino Cold Dark Matter}\\

In \cite{theorem,susyseesaw}
it has been shown that $L$ violating sneutrinos give rise to loops
containing neutralinos and (light) sneutrinos which contribute to the 
(Majorana) mass of the neutrino. In \cite{sndb} a scan over a wide range
of the SUSY parameter has been carried out and ``average'' upper
limits 
\be \label{neutrinobound}
\Delta m (e) < 8 MeV \ \ , \ \ 
\Delta m (\mu) < 6 GeV
\ee
have been deduced (for neutrino mass limits $m(e)<15eV$ and 
$m(\mu)<170keV)$ while the third generation mass-splitting remains 
virtually unconstrained. The current bound on the effective Majorana 
neutrino mass $\langle m_{ee} \rangle = \sum_i' U_{ei}^2 m_i < 0.36 eV$
(90$\%$ $c$.$l$.) from the 
Heidelberg-Moscow experiment \cite{0nexp} yields in the case
$U_{ei} \approx \delta_{ei}$ the relation 
$\Delta m (e) < 13  keV$ for an average sneutrino mass of $100 GeV$. 
Direct contributions to $\0n$ have been considered in \cite{sndb} and 
the resulting limits on $\Delta m (e)$ are less stringent. 
Therefore the limits from baryogenesis are considerably less severe
than the ones from neutrino masses in the first generation case for
plausible ranges of $T_C$ and superpartner masses. 
For the second generation the limit is comparable or slightly 
more stringent for values of $T_C$ not much smaller than $150GeV$
and superpartner masses not bigger than $500 GeV$.
In the third generation case for plausible values of
$T_C$ baryogenesis yields more stringent limits than neutrino
masses do. 

The constraints on the sneutrino mass-splitting are especially
interesting for the hypothesis of sneutrino CDM. In \cite{cdm} it
has been shown that a sneutrino with mass of around $70 GeV$ may
be a viable CDM candidate if the mass-splitting exceeds about
$5GeV$ (so that $Z^0$-mediated $s$-channel coannihilation of the
LSP sneutrino with its heavier partner is sufficiently suppressed)
and $M_1$ is bigger than about $200 GeV$ (so that $L$-conserving
LSP-sneutrino pair annihilation is sufficiently suppressed). Therefore the 
low energy limits eq. (\ref{neutrinobound}) imply that the light 
third generation sneutrino is a viable CDM candidate. 
On the other hand, in case {\bf I)} discussed in the last chapter 
the results displayed in figure \ref{figresult} (solid line) imply 
that sneutrino CDM is firmly ruled out for plausible values of the
critical temperature $T_C \gsim 150 GeV$. 

The same conclusion is valid in scenario {\bf II)} for plausible
values of $T_C$ if $M_1$ and $M_2$ bear the same sign as long as
the gauginos are not exceptionally heavy ($M_1 \gsim 1TeV$). 
However, in case {\bf II)} the bound on the sneutrino mass-splitting 
could become sufficiently large to allow for sneutrino CDM if $M_1$ and
$M_2$ are of opposite sign. For plausible values of $T_C \gsim 150 GeV$
the ratio $M_2/M_1=-3\alpha_W/5\alpha_Y$ ($M_2/M_1=-\alpha_W/\alpha_Y$) allows
for a sneutrino mass-splitting being large enough to be compatible with
the sneutrino CDM hypothesis as long as $M_1$ is bigger than about 
$500 GeV$ ($300 GeV$).
In particular, this statement is compatible with the lower limit 
on $M_1$ from the requirement that the sneutrino relic abundance 
should be sufficiently high to account for the CDM. 

\vspace{0.5cm}
\noindent
{\bf VI \hspace{0.2cm} Summary}\\

In conclusion, in scenarios with $L$ violation in the light sneutrino 
sector it has to be made sure that sneutrino-induced interactions which
take place in the early universe at temperatures below the electroweak 
symmetry breaking scale do not erase the BAU generated at some higher 
temperature. This may happen if the electroweak phase transition is
second or weakly first order, so that sphaleron-induced processes are
still operative at temperatures below the critical temperature $T_C$.
The $L$-violating interactions considered here result in  
constraints on the mass-splitting of the light sneutrino mass states 
which are less stringent than the ones which can be derived from the 
contribution of $L$-violating sneutrinos to neutrino masses for the
first generation. For the second generation the limits are more
stringent for plausible values of $T_C \gsim 150 GeV$ and for 
superpartner masses smaller than $500 GeV$ except for the case of
an almost complete cancellation of the bino and wino contributions
for a sneutrino LSP of mass ${\cal O}(100GeV)$. 

In particular, the constraint on the third generation mass-splitting 
is of interest for the sneutrino Cold Dark Matter hypothesis. Low energy
limits on the
sneutrino mass-splitting suggest that the lighter of the third
generation sneutrino states with mass of around $70 GeV$ may serve 
as a CDM candidate if $M_1 \gsim 200 GeV$. This hypothesis is not
compatible with the observed Baryon asymmetry if the gauginos or further
sleptons other than the sneutrino are still present in the plasma
at temperatures below $T_C$ or if the masses associated to the bino
and wino bear the same sign. On the other hand, a scenario with
opposite sign gaugino mass parameters and both gauginos sufficiently 
heavy $M_1,M_2 \gsim 500 GeV$ (and slepton masses
bigger than $T_C$) is compatible with the sneutrino CDM hypothesis.\\   


\centerline{\bf Acknowledgments}

\vspace{0.5cm}
We thank D. Semikoz and U. Sarkar for helpful discussions. St. K. would 
like to thank the Institute for Nuclear Research in Moscow for hospitality
and the Deutscher Akademischer Austauschdienst for financial support.

\end{document}